\documentclass[aps,preprint]{revtex4-1}

\usepackage[breaklinks=true,colorlinks,citecolor=blue,linkcolor=blue,urlcolor=blue]{hyperref}

\usepackage{graphicx}
\usepackage[normalem]{ulem}
\usepackage{color}
\usepackage{amsmath}
\usepackage{enumitem}
\usepackage{amsthm}
\usepackage{hyper ref}

\begin{document}
\title{Quantum and classical frontiers of noise}

\author{Xavier Oriols}
\email{xavier.oriols@uab.cat}
\affiliation{Departament d'Enginyeria Electr\`onica, Universitat Aut\`onoma de Barcelona, 08193-Bellaterra (Barcelona), Spain}

\begin{abstract}
This paper is an introduction to the eleven works of the special issue on \emph{Quantum and Classical Frontiers of Noise}. The weather, and its butterfly effect, is the typical example that explain why many natural phenomena are, in fact, not predictable with certainty. Noise in classical or quantum phenomena, understood as the difference between the empirical output of an experiment and its statistical prediction, is a measure of such uncertainty. One of the great contributions of noise appeared in 1905 when Einstein showed how the noisy (Brownian) motion of a dust particle seen in a microscope provided uncontroversial evidences on the existence of the (unseen) atoms. The aim of this special issue is to provoke discussions on how noise can help us to better understand (the reality behind) classical and quantum theories, and the frontier between them.
\end{abstract}

\maketitle


\section{How much of our world is predictable with certainty?}
\label{sec1}

Albert Einstein, in the paper Physics and reality  \cite{einstein}, pointed out the possibility of living in a bizarre world without comprehensible explanations for natural phenomena. He wrote: \emph{``The fact that [the world] is comprehensible is a miracle"}. Similarly, Eugene Wigner wrote: \emph{``The unreasonable effectiveness of mathematics in the natural science ....is a wonderful gift which we neither understand nor deserve"}  \cite{wigner}. Both reflections were inspired by the previous work of the German philosopher Immanuel Kant who wrote the very same idea almost two centuries before: \emph{``The eternal mystery of the world is its comprehensibility."}. Fortunately, it seems that we  live in a comprehensible world. However, if our world is comprehensible, how much of it can we predict with certainty?

After the work of Isaac Newton in 1687, it was believed at that time that all motions in heaven and earth were predictable with certainty and, in fact, explainable with just three simple mathematical laws  \cite{newton}. Kant was deeply impressed by Newton's work and he wanted to understand its limitations. Kant divided scientific knowledge into three parts: appearance, reality and theory. Appearance is the content of our sensory experience of natural phenomena, i.e. the empirical outcome of an experiment. Reality is what lies behind all natural phenomena. A theory is a human model that tries to mirror both appearance and reality. A useful theory might predict the outcome of an experiment in a laboratory or the observation of a phenomenon in Nature. The predictions of a theory can be certain \cite{footnote1} or statistical  \cite{footnote2}. Empiricists believe on experimental outcomes (what Kant called appearance) and refuses to speculate about a deeper reality. On the other hand, realists believe that good physical theories explain, or at least provide clues about, the reality of our comprehensible world.

As all human creations, there are successful and unsuccessful theories. When in 1864 James Clerk Maxwell conjectured that light was an electromagnetic vibration, it was believed that all waves had to vibrate in some medium. The medium in which light presumably travels was named  \emph{luminiferous ether}. During almost a century eminent scientists believed blindly on that concept. Nowadays, the \emph{luminiferous ether} plays no role at all in modern physical theories \cite{herbert}. The atomicity of matter is an example of a very successful theory. It was introduced by the British chemist John Dalton in 1808 to explain why some chemical substances need to combine in some fixed ratios. During one century it was thought that atoms was a crazy idea. Marcelin Berthelot said \emph{``who [has] ever seen a gas molecule  or an atom?''}, expressing the disdain that many chemists felt for the unseen atoms, that were inaccessible to experiments  \cite{herbert}. Even its defenders saw little hope of ever directly verifying the atomic hypothesis. Nowadays, the fact that everything is made of atoms is one of the most precious knowledge that we get on how Nature works \cite{feynmann},  and their images  are even routinely seen in the screens of scanning tunneling microscope apparatuses  \cite{binning}.   

At this point, it is important to think about the possible relation between a theory and its predictions \cite{wiseman}. Surprisingly, a deterministic theory can provide statistical predictions of the outcomes, while a stochastic theory may provide predictions without uncertainty. We say that a natural phenomenon is predictable with certainty when, for a given initial preparation of the experiment, we can know from the theory if a particular outcome will occur or not, without ambiguity. We refer to a theory as a deterministic one when the specification of the complete  \cite{footnote3} state of a system at one time, together with the laws of the theory, is compatible with only one state of the system at any other time.

In order to clarify the previous discussion, let us repeat an example found in one the papers of this special issue  \cite{oldofredi}. We assume an ideal gas in a bottle governed by classical Newtonian mechanics. Although, according to the classical theory, the positions and velocities of all particles of the gas have well-defined values, it is impossible for us to know all the $10^{23}$ values. With the help of statistical mechanics, we can find which initial conditions are typical and which atypical and predict a statistical distribution of the total kinetic energy of the gas, but not its exact value at one particular time. The exact value of the kinetic energy of the gas, strictly speaking, cannot be predicted with certainty. It is very important to realize that the fact that the theory has well-defined values does not imply that we, the humans, know these values in an experiment. 

There are other theories, for example the well-known orthodox (Copenhagen) quantum theory that are not deterministic. Even if the wave function has well-defined initial values, in general, the theory itself does not determine with certainty the occurrence or not of an (observed) outcome. According to the theory, most of the times that a measurement is done, a stochastic perturbation acts on the wave function. This is the so-called \emph{collapse} of the wave function. Therefore, the results of such experiments cannot be predicted with certainty, and only statistical predictions are available. 

The orthodox (Copenhagen) non-deterministic quantum theory can in fact predict with certainty some particular outcomes. According to the equations of motion of this theory, the state after a (projective) measurement is an eigenstate of the operator associated to the measurement. Therefore, after a second measurement (assumed to be done instantaneously after the first one to avoid unnecessary complications due to the evolution of the system) with the same (projective) measurement, we can predict with certainty that the outcome of the second measurement will be the same as the first one  \cite{landau}. 

From the very beginning, the correct modeling of the measurement of quantum systems has been recognized as one of the the most puzzling and intriguing problems. It has direct implications on how quantum uncertainty is understood and on how the quantum-to-classical transition is explained. The so-called \emph{measurement problem}  \cite{bell,herbert,maudlin} appears because the use of a wave function governed by a linear dynamical equation of motion (like the unitary Schr\"odinger evolution) allows for several different outcomes (the Schr\"odinger's cat, dead and alive, is the typical example). Then, if we assume that in our macroscopic world, when a measurement is done, only one outcome occurs (the cat alive or the cat dead, but not both), we have a problem because there is noting in the wave function for favor one or the other outcomes. One possible solution to the problem is arguing that we have one world where the cat is alive and another different world where  the cat is dead. This is the many worlds interpretation discussed in the paper \cite{barret} of this special issue. Another solution to the measurement problem consist in assuming that there is only one world, but that the wave function alone does not provide a \emph{complete} specification of the quantum state. For example, as done in the paper \cite{sole} of this special issue, we can add a trajectory for each particle of the quantum system. Bohmian mechanics is the theory behind such account. Finally, another solution is just arguing that the mentioned unitary Schr\"odinger-like evolution of the wave function is not always applicable, and that new quantum non-linear equations are required. This is dealt with in the papers  \cite{ghirardi} and  \cite{vacchini} of this special issue. The former is the GRW interpretation that complements the Schr\"odinger evolution with a new stochastic term in the equation of motion of the wave function, while the latter is the orthodox (Copenhagen) interpretation. This last explanation imposes that  the Schr\"odinger equation has to be substituted by a new law, known as the collapse law \cite{landau}, every time that a measurement is done. Certainly, in spite of being not free from controversies, this last option seems the most widely accepted solution to the measurement problem. An eminent physicist John S. Bell, which became well-known for his inequalities that have the implications that our world is non-local, however, in 1982  \cite{bell}: \emph{``The concept of measurement becomes so fuzzy on reflection that it is quite surprising to have it appearing in physical theory at the most fundamental level."}.

I conclude this part providing an answer to the question posted in the title of this section: How much of our world is predictable with certainty? Accepting a coarse-grained computation of the possible outcomes, many macroscopic phenomena seem to be predictable with (a coarse-grained) certainty. On the contrary, we realize that most Natural phenomena, or outcomes in laboratories, cannot be predicted with absolute precision. Contrarily to Newton's thoughts, we can compute the trajectory of (the center of mass of) Jupiter with great precision, but we cannot predict the exact size of its great red spot (a persistent anticyclonic storm at south of Jupiter's equator). 

\section{Signal and noise}
\label{sec2}
In this section, I provide an informal relation between uncertainty and noise. We have already mentioned several classical and quantum examples of experiments whose possible $N$ outcomes $a_1,a_2,a_3,...,a_N$ are not predictable with certainty. We have only access to the probability of the different outcomes $a_i$ defined through the function $P(a_i)$. From this statistical information, we can estimate the mean value:
\begin{eqnarray}
\langle A \rangle= \sum_{i=1}^N P(a_i)a_i.
\end{eqnarray} 
The knowledge of $\langle A \rangle$ does not allow us to predict with certainty which outcome $a_i$ will occur in a given experiment. Nevertheless, we are sure that the outcomes in a set of identical experiments will satisfy $\langle A \rangle$ when properly averaged. The value $\langle A \rangle$ is a statistical prediction. It seems very reasonable to talk about $\langle A \rangle$ as the signal of our experiment. Once we have a precise definition of what is a signal, the meaning of the noise is just the difference between the measured (observed) outcome $a_i$ and the signal. Technically, since we have to average over $N$ possible outcomes, in order to avoid positive and negative cancellations of the difference, it is better to quantify the noise as:
\begin{eqnarray}
\sigma^2=\sum_{i=1}^N (a_i-\langle A \rangle)^2 P(a_i),
\end{eqnarray} 
where $\sigma$ is the so-called standard deviation. For a set of identical experiments where the outputs are always $a_1$, the mean value is $\langle A \rangle=a_1$ and there is no noise, i.e. $\sigma=0$. Noise is, in this sense, a measure of the uncertainty of the classical or quantum experiments. 

Noise can have many different origins. For example, a spurious  distortion in the measurement apparatus between what it really measures on the system and what is indicated in the pointer. Other sources of noise can be attributed to our ignorance about the (microscopic) initial conditions of a complex system or to a fundamental stochastic element in the theory. Sometimes, we can reduce the noise of a particular phenomena by improving the technological tools that allows us to better determine the initial conditions of the experiment. Others, it is the theory itself that does not allow us to better predict the outcomes  \cite{acin}.

Noise in classical or quantum phenomena cannot be fully domesticated. It is not only a hindrance to signal detection, but it is indeed a valuable source of information (not present in the signal) that help us to get a deeper understanding on how Nature works. Perhaps, the first great contribution of noise into our understanding of Nature was the demonstration of the real existence of atoms  \cite{cohen}. As mentioned in section \ref{sec1}, during the 19th century there were an intense debate about the reality of atoms  \cite{herbert}. The success of thermodynamics convinced many scientist that atoms and molecules were imaginative fictions and that the real underlying component of the Universe was energy, in its various forms. In 1905, the same year that Einstein conceived the theory of relativity that demolished the \emph{luminiferous ether}, he published a paper on Brownian motion that pointed the way to conclusive experiments bearing on the existence of atoms  \cite{einstein2}. Whenever micron-sized particles are suspended in a liquid they undergo a perpetual quivering dance whose origin had been a mystery since its discovery in 1828 by the Scottish botanist Robert Brown  \cite{brown}. Einstein showed that although the number of atoms striking the Brownian particle from each direction is equal on the average, the noise on the number of atoms (the fluctuations of this number with respect to the average value) leads to unbalanced forces in random directions. For a very large particle, the pressure of the surrounding atoms is indeed evenly balanced, but for a small particle, the fluctuations in the number of atoms impinging the particle is enough to propel it in an unpredictable direction with a predictable force, as indicated in the conclusions of the paper  \cite{vacchini} of this special issue: Can discussions on noise allow a deeper understanding of the quantum reality?

\section{The frontier between classical and quantum mechanics}
\label{sec3}

I started this paper by wondering about the comprehensibility of our world  \cite{einstein,wigner}. In any case, our comprehension of how Nature works is at the present time partitioned in several theories. Each theory describes reality in a different manner. We have been unable yet to develop a theory of everything, a final and grand unified theory. Classical mechanics provides good predictions with a (coarse-grained) certainty for macroscopic objects when particle velocities $v$ are much slower than the speed of light $c$. The classical theory visualizes the reality in terms of point-particles with well-defined trajectories. However, the theory completely fails for larger velocities or smaller lengths. For $v \approx c$, classical mechanics is substituted by relativity. The frontier between classical mechanics and relativity is well understood. The equations of motion of the theory of relativity depend on the ratio $v/c$ and in the limit $v/c \rightarrow 0$ the relativistic equations of motion become classical. 

For small (microscopic) lengths, the classical theory has to be substituted by quantum ones. However, the quantum-to-classical frontier is still unclear, diffuse. It is widely assumed that the equations of motion of the quantum theories must lead to the classical ones in the \emph{proper limits}. However, we have not yet a clear way to understand what \emph{proper limits}  means.  Since the beginning of the quantum theory a century ago, the study of this frontier has been a constant topic of debate. The first attempts to reach classical dynamics from quantum mechanics revolved around simple assumptions like a Planck's constant equal to zero \cite{landau}, large quantum numbers  \cite{bohr20}, large mass  \cite{angel}, etc. All these arguments, in spite of being useful in some simple experiments, are clearly insufficient to get a clear and rigorous understanding of the quantum-to-classical transition  \cite{Ballentine90,Mako06,klein12}. An evidence of the difficulties in understanding what the \emph{proper limits} means is that, although most macroscopic objects behaves classically, macroscopic quantum superposition is still present in some systems and accessible experimentally  \cite{oriol}. 

During last decades, decoherence has been invoked as an empirical explanation of the quantum-to-classical transitions \cite{camilleri}. However, one has to worry that decoherence cannot be invoked as a type of new postulate that forces quantum phenomena to become classical. On the contrary, decoherence should appear as a consequence of the postulates of the theory itself. For example, within the orthodox (Copenhagen) interpretation, decoherence has been defined for an open quantum system as a result of the entanglement between its degrees of freedom and those of the environment. Decoherence explains then why the reduced density matrix of an open system becomes diagonal, but many scientist argue that it still does not solve the quantum-to-classical transition satisfactorily because one has to invoke afterwards  the intriguing collapse law to select only one element of the diagonal of the density matrix \cite{Zurek03,Giulini96,schlosshauer14}. Other interpretations of quantum phenomena, like the ones discussed in this special issue can shed light on the problem of decoherence from different points of view.  

The fact that the quantum-to-classical transition is still so badly understood at a fundamental level is even more surprising because many of the new spectacular scientific and technological developments are nowadays playing in the nanotechnology arena. In spite of our difficulties in properly understanding this frontier, this new developments involve indeed this macroscopic-microscopic (quantum-to-classical) frontier.

\section {Outlook of the present special issue}

The aim of the eleven papers of this special issue is to highlight the power of noise (or uncertainty) as a useful practical and theoretical tool to better comprehend how Nature works. The papers, written by experts in philosophy, physics and engineering are grouped into three different subsections. Next, I introduce each subsection and briefly contextualize the different works. 
 
\subsection{Fundamental noise in quantum phenomena}
 \label{grup1}

As I mentioned, some sources of noise, for example those related to our ignorance of some details of the environment, are common in classical and quantum phenomena. However, quantum phenomena show an additional source of noise not present in classical ones. If we measure the same property in a set of quantum systems, all of them identically prepared, the measured values do not generally coincide.  We can only predict the statistical distribution of such values. There is absolute uncertainty in most quantum experiments. The different ontological explanations of this source of noise are discussed in this section. The \emph{measurement problem} mentioned in section \ref{sec1} is very present here. Different interpretations (inspired by different solutions of the measurement problem) explain uncertainty assuming very different realities  \cite{everett,ghirardi,bohm}.\\

The first paper \textbf{Typicality in Pure Wave Mechanics} was written by Jeffrey Barrett  from the  University of California in Irvine, United States  \cite{barret}. It explains how a consistent interpretation of the quantum uncertainty can be conceived assuming that the wave function is a complete description of a quantum state and without the controversial collapse law. Such interpretation is known in the literature as the \emph{Many worlds} interpretation. The origin of the \emph{Many worlds} interpretation was the relative state formulation done by Hugh Everett \cite{everett}. Indicating that his theory was fully deterministic, Everett originally entitled his thesis ``Wave Mechanics without probability''. Later, Everett's work was further developed by Bryce DeWitt and co-workers  \cite{witt} and renamed as many-worlds. The wave function, interpreted by Born as a mathematical entity that encapsulates all different probabilities in a unique world, is reinterpreted here in a quite different way. Every time that the wave function dictates different possible outputs, the world splits into different worlds. For example, in the famous Schrodinger's cat, after the measurement of the state of the cat, the cat is alive in one world and dead in another. Each world is as \emph{real} as the one where I am writing this introduction. The world where I am writing corresponds to one sequence of the possible outputs. The puzzle is to explain the standard quantum statistics when the theory is fully deterministic and every possible measurement outcome in fact occurs. In short, Everett’s goal was to show that the standard quantum probabilities would appear to be valid in a \emph{typical} world. \\

The second paper entitled \textbf{How does Quantum Uncertainty Emerge from Deterministic Bohmian Mechanics?} was written by Albert Sol\'{e} from Universitat de Barcelona and Xavier Oriols from Universitat Aut\`{o}noma de Barcelona, both from Spain, Damiano Marian from Universit\`a di Pisa in Italy and Nino Zangh\`{i} from Universit\`a di Genova and the INFN sezione di Genova, Italy  \cite{sole}. The paper explains another possible interpretation of quantum mechanics with deterministic laws and without the stochastic collapse law. This interpretation is usually named Bohmian mechanics in honour of David Bohm \cite{bohm} or de Broglie-Bohm theory to acknowledge also the initial work of Louis de Broglie  \cite{Oriols,ABM,durr}. In this theory, the wave function is not a complete description of a quantum state and, apart from the wave function, the actual position of the particles of the system are required. The whole state of the system (wave and particles) evolves deterministically. Thus, the Bohmian state can be compared with the state in classical mechanics, which is given by the positions and momenta of all the particles, and which also evolves deterministically. However, in Bohmian mechanics the initial position cannot be measured without disturbing the wave function. Even worst, the initial wave function cannot be measured with an apparatus, it can only be prepared). Since we ignore the initial conditions of a system, the outcomes of a quantum experiment become uncertain for us even when using a deterministic theory as Bohmian mechanics. \\

 The third paper \textbf{Conceptual Difficulties of standard quantum mechanics and the role of noise in overcoming them} was written by GianCarlo Ghirardi from the University of Trieste and the Abdus Salam International Centre for Theoretical Physics, in Trieste, Italy  \cite{ghirardi1}.  As I mentioned, the stochastic collapse law of the standard (orthodox) quantum mechanics is seen for some people as a suspicious/questionable way of solving the measurement problem. It does not explain when the measurement takes place (at the beginning of the experiment? at the end? when somebody -with a PhD- is looking at the system  \cite{bell}, ...). In a real experiment, what is the measured system and what is the measurement apparatus? One way of solving these ambiguities of the orthodox theory is by arguing that a collapse-like phenomena happens in our world in a \emph{natural} way, without depending on the (human as observers) measurement. This theory, known as the GRW  \cite{ghirardi} to acknowledge the authors Giancarlo Ghirardi, Alberto Rimini, and Tullio Weber, has the additional advantage that provides a trivial understanding on why our microscopic world satisfies a linear principle, while the macroscopic world does not. The equations of motion of the theory includes, apart from a Schr\"odinger-like linear evolution, an additional stochastic evolution (noise) that provides a \emph{natural} collapse and accounts for quantum uncertainty. \\

The last paper of this section is entitled \textbf{Quantum Noise from Reduced Dynamics} and it was written by Bassano Vacchini from Universit\`a degli Studi di Milano, in Italy  \cite{vacchini}. Many scientists affirm that the solutions mentioned above to solve the measurement problem (and the underlying reality that these solutions postulates) are even more unsatisfactory than the collapse law itself. What is objectively true is that almost all important developments of quantum mechanics in the last century have been developed within the collapse law of the orthodox theory. In particular, in this paper a review of the widely used orthodox techniques to deal with open systems, interacting with the environment, is presented. Averaging over the environmental degrees of freedom leads to stochastic quantum equations for the subsystems. Such equations include the constraints arising from the statistical structure of quantum mechanics (the collapse law) and the uncertainties appearing due to the fact that the system is open (without a complete knowledge of the environment). Simple examples are used to didactically explain the dynamics of such open systems. As indicated in the conclusions of this paper   \cite{vacchini}, perhaps, a better understanding of the role of noise in some quantum phenomena will allow us to discern between different interpretations. One of the main goals of this special issue is motivating research towards this direction.

\subsection{Differences of noise in classical and quantum phenomena}
 \label{grup2}
 
The quantum or classical equations of motion are different. Therefore, there are quantum features of noise phenomena without classical counterpart. For example, among many others, the Leggett-Garg inequalities  \cite{legget} show how a quantum measurement itself has strong implications on the noise. Bell's inequalities  \cite{belline} have been further developed to show how faster-than-light causation has also effects on quantum noise. Quantum and classical chaos become very close at the experimental side, but not in their theoretical description. In this section, some examples are discussed. \\

The paper \textbf{Sequential measurement of displacement and conduction currents in electronic devices} was written by Guillermo Albareda from University of Barcelona in Spain, Fabio Lorenzo Traversa from  University of California, San Diego in United States  and Abdelilah Benali from Universitat Aut\`{o}noma de Barcelona in Spain  \cite{albareda}. It deals with the measurement of the electrical current in quantum devices at very high frequency. From the extension of the Ramo-Schockley-Pellegrini theorem to quantum devices, the authors define a positive-operator valued measure for the total (conduction plus displacement) electrical current.  They have implemented the resulting measurement protocol into a quantum electron transport simulator and the computed results for a resonant tunneling diode depends greatly on the details of the implementation of the total current operator \cite{albareda2,albareda3}. In this sense, the paper can be considered as a practical example of the difficulties of the measurement mentioned in the well-known Leggett-Garg inequalities \cite{legget}.\\

The paper \textbf{Origin of Shot Noise in Mesoscopic Cavities} was written by Massimo Macucci and Paolo Marconcini, from Universit\`a di Pisa in Italy  \cite{macucci}. The paper studies the shot noise suppression in mesoscopic cavities. It starts by discussing the two origins of (classical or quantum) shot noise: the stochastic injection of electrons (because we are dealing with an open system) and its stochastic transmission  (due to the quantum noise discussed in section \ref{grup1} or to other classical sources of stochastic dynamics inside the structure). After reviewing the relevant literature, the authors present some numerical results to conclude that shot noise behaviour in mesoscopic cavities can be explained without any need for classical chaotic dynamics and only on the basis of quantum chaos resulting from diffraction at the constrictions. The paper does also suggest several reflections on the relation between classical and quantum chaos. Can a better understanding on the origin of chaos improve our understanding of the frontier between quantum and classical theories? \\

The paper \textbf{From the universe to subsystems: Why quantum mechanics appears more stochastic than classical mechanics} was written by Andrea Oldofredi, Dustin Lazarovici and  Michael Esfeld from Universite de Lausanne in Switzerland, and Dirk-Andre Deckert from Ludwig-Maximilians-Universit\"{a}t M\"{u}nchen, in Germany  \cite{oldofredi}. This paper emphasizes through didactic examples that an equation of motion, without fixing the initial conditions, become useless for making exact predictions. What happens when the number of particles of an ideal gas in a box is so large ($10^{23}$) that it is impossible to determine the initial conditions of all the particles? In such systems, the concept of typical and atypical initial conditions become very relevant. A (coarse-grained) uniform distribution of particles is typical, while all particles concentrated together in a small region of the box is atypical. From the concept of typicality, we can easily move to the concept of probability for either classical or quantum phenomena. Boltzmann was the first to develop such strategy in classical mechanics. It is very relevant to emphasize that, as Einstein already noted, Boltzmann's procedure to build probabilities in classical mechanics comes from outside of the equations of motion of the theory itself. By means of the examples of both classical and Bohmian mechanics, the authors explain what quantum and classical probabilities have in common and where exactly their differences lie.\\

The paper \textbf{Randomness and non-locality} was written by Gabriel Ignacio Senno, Ariel Bendersky, and Santiago Figueira from Universidad de Buenos Aires in Argentina  \cite{senno}. The paper deals with the relationship between uncertainty and non-locality. The authors provide a summary of recent advances in the interplay between these two important concepts from a quantum information point of view.  According to the orthodox (standard) quantum theory, the collapse of the wave function is the fundamental source of uncertainty in the quantum regime. The important point here is that a wave function of $N$ particles \emph{lives} in a configuration space of $3N$ dimensions (not in the three-dimensional physical space). Therefore, the stochastic perturbation of the wave function in the configuration space due to the collapse after a measurement, implies a faster-than-light interaction among particles. This effect, implicit in the Bell's inequalities  \cite{belline}, implies a deep connection between quantum non-locality and quantum uncertainty.\\

\subsection{Noise in the quantum-to-classical frontier}
 \label{grup2}
 
Our everyday experience certifies that our macroscopic world follows, most of the time, classical laws, while the microscopic (atomistic) world follows quantum mechanical ones. Surprisingly, as discussed in section \ref{sec3}, the transition between both theories is not at all obvious. Discussions of some systems at this frontiers are presented below.\\

The first paper \textbf{On the Classical Schr\"odinger equation} was written by Albert Benseny from Okinawa Institute of Science and Technology Graduate University in Japan, David Tena and Xavier Oriols  both from Universitat Aut\`{o}noma de Barcelona in Spain  \cite{benseny}. The paper discusses the classical Schr\"odinger equation. The authors emphasize how a (usually unnoticed) univaluedness condition on the Hamilton's principal function is assumed during the transition from Newton mechanics to the classical Schr\"odinger equation. This additional assumption implies that the trajectories described by the classical Schr\"odinger equation do not cross in the configuration space, providing unexpected non-classical features. The authors also explain the strategy of reaching the classical wave function by departing from quantum Bohmian mechanics and invoking the quantum-to-classical transition when a large number of identical particles are invoked. Here, the center of mass plays the role of coarse-grained outcome mentioned along the introduction. The authors also show that a condition of dealing with a very sharp wave packet of the center of mass is mandatory, which implicitly avoids the non-classical features mentioned above. The paper is an example of the difficulties discussed in Section \ref{sec3} in the quantum-to-classical transition: Why does the quantum wave function seems to become irrelevant in classical scenarios?\\

The paper \textbf{Noise in quantum devices: A Unified Computational Approach For Different Scattering Mechanisms} was written by Damiano Marian from Universita di Pisa, Italy and Enrique Colom\'{e}s from Universitat Aut\`{o}noma de Barcelona in Spain \cite{marian}. From the point of view of simulation, an electronic device is an extraordinary complex system because it is a many-body open system out of (thermodynamic) equilibrium. The typical strategy to simulate such complex systems is to discern between the relevant processes that determine the global behavior of the device and those processes that have a smaller effect on the device. Then, we only model the equations of motion of the relevant processes and the other processes are just introduced later as perturbation. The author use the notion of conditional wave function, the wave function of a subsystem in Bohmian mechanics, to define an effective equation of motion for electrons in a subsystem. Then, through the use of different degrees of approximations, transport is modeled in a simple tunneling barrier as well as in a electronic system interacting with a bath of phonons \cite{marian2}. The paper also debates on the risk of making excellent mathematical approximations in the equations of motion, without any physical justification apart from fitting experimental results. In the literature, there are many models for electronic transport that allow a proper interpretation of classical and quantum devices and the transition between them (by properly fitting the parameters of the scattering rates included in the equation of motion). This has the risk of producing unfalsifiable theories  \cite{camilleri,oriols2}. \\

Finally, the paper \textbf{Zero Thermal Noise in resistors at zero temperatures} was written by Laszlo B Kish from Texas A\&M University in United States, Gunnar A Niklasson and Claes-Goran Granqvist from Uppsala University in Sweden \cite{kish}. The paper points out interesting paradoxes that appears in typical electronic devices when analyzing noise at zero temperature. The fluctuation-dissipation theorem for the voltage and current present in a resistance shows that the noise can be modeled as the sum of a classical plus a quantum (zero point) term. The classical term vanishes when the temperature approaches zero, while the quantum term still predicts non-zero noise voltage and current at zero temperature. The authors discuss the paradoxes, possible theoretical flaws and experimental artifacts of getting noise at zero temperature in practical experimental situations and they point out a possible solution. They also insist that its resolution is essential to determine the real error rate and fundamental energy dissipation limits of the state-of-the-art logic gates. Can such paradoxes shed light on a better understanding of the quantum-to-classical transition? \\

\section {Final remarks}

The investigation of noise is important because it can reveal information on classical and quantum phenomena not available from mean values. Noise is not only a hindrance to signal detection, but it can help us to enlighten our understanding on how Nature works. Certainly, the investigation of noise is a more difficult task than the investigation of mean values, either from a experimental or theoretical point of view (in part because of the still puzzling quantum measurement problem). 

Everyday our society is impressed by spectacular scientific and technological advances in nanotechnology. Let me mention few of them. We are now able to measure the trajectories of quantum particles (photons) through weak values \cite{photons}. With weak measurements, we have experimental control on how much distortion (i.e. how much \emph{collapse}) is produced on the measured state \cite{weak}. We get macroscopic superpositions of quantum objects in laboratories \cite{oriol}. Quantum (analog) simulators allow us to create artificial atoms with artificial interactions \cite{cold}. In spite of such unquestionable control over quantum systems, many fundamental questions still remain unanswered.  An improved comprehension of how this nano world works is very relevant, not only from a fundamental point of view, but also from a practical point of view to be able to exploit all the great (quantum engineering) possibilities that seems to be hidden ``at the bottom''.  

Inspired by one of first great contributions of noise in understanding the atomicity of matter through the noisy (Brownian) motion of a dust particle  \cite{einstein2}, the aim of this special issue, written by different experts ranging from philosophy to applied physics, is to point out that noise can be a useful tool to provide a better comprehension of the reality behind classical and quantum theories, and the frontier between them. 

\section *{Acknowledgments}

I sincerely thank all the authors of this special issue for enlightening discussions. This work has been partially supported by the Fondo Europeo de Desarrollo Regional (FEDER) and Ministerio de Econom\'ia y Competitividad through the Spanish Projects No. TEC2012-31330 and No. TEC2015-67462-C2-1-R and the Generalitat de Catalunya (2014 SGR-384). This project has received funding from the European Union's Horizon 2020 research and innovation programme under grant agreement No 696656.


\begin{thebibliography}{99}
\bibitem{footnote1}
The prediction is certain when the theory provides one value when forecasting what would happen in the outcome of an experiment (the definition assumes that the theory agrees with experiments)
\bibitem{footnote2}
The prediction is statistical when the theory itself gives a list of values, each one with its probability, when explaining the possible value of the outcome in an experiment. When repeating the experiment many times, with identical preparation, the different values of the outcome satisfy the predicted probabilities.
\bibitem{footnote3}
The meaning of complete specification of the state is defined by the theory itself.
\end{thebibliography}
\end{document}